\documentclass[prb,twocolumn,showpacs,superscriptaddress]{revtex4}

\usepackage{amssymb,amsmath,graphicx}

\begin{document}

\title{Geometric strong segregation theory for compositionally
  asymmetric diblock copolymer melts}

\author{C. B. Muratov}

\affiliation{Department of Mathematical Sciences, New Jersey Institute
  of Technology, Newark, NJ 07102, U.S.A.}

\author{M. Novaga}

\affiliation{Dipartimento di Matematica, Universit\`a di Pisa, Largo
  B. Pontecorvo 5, Pisa 56127, Italy}

\author{G. Orlandi}

\affiliation{Dipartimento di Informatica, Universit\`a di Verona,
  Strada le Grazie, 15, Verona 37134, Italy}

\author{C. J. Garc\'ia-Cervera}

\affiliation{Department of Mathematics, University of California, 
Santa Barbara, CA 93106, U.S.A.}

\date{\today}

\begin{abstract}
  We have identified the effect of the Wigner-Seitz cell geometry in
  the strong segregation limit of diblock copolymer melts with strong
  composition asymmetry. A variational problem is proposed describing
  the distortions of the chain paths due to the geometric constraints
  imposed by the cell shape. We computed the geometric excess energies
  for cylindrical phases arranged into hexagonal, square, and
  triangular lattices and explicitly demonstrated that the hexagonal
  lattice has the lowest energy for a fixed cell area.
\end{abstract}


\maketitle

Block copolymers are a well-known class of smart materials that can
produce a wide variety of complex equilibrium microstructures
\cite{bates99,goodman,aggarwal,bates90}. Since the late 70's, these
systems have received significant attention by theorists, and many of
the copolymer system phases are now well explained on the basis of
energy minimization arguments
\cite{helfand75,helfand76,helfand78,helfand80,leibler80,semenov85,%
  ohta86,likhtman94,olmsted98}. Further advances in computational
techniques allowed studies of the phase behavior in block copolymer
systems with the help of direct numerical solution of models which
explicitly incorporate statistical mechanics of polymer chains,
providing a connection between the observed microstructures and the
underlying microscopic material parameters
\cite{matsen94,matsen96jcp,groot98,murat99,matsen01,matsen02,%
  schultz02,cochran06,bosse06}. More recently, a renewed interest in
block copolymer systems was stimulated by the studies of geometrically
constrained systems, such as block copolymers confined to the surface
of a substrate or the inside of nanopores
\cite{lambooy94,knoll02,xiang04}. Numerical studies of these systems
showed a rich variety of microstructures with very intricate
geometries \cite{li06,chen07,chantawansri07,yu07}.

An important regime in which the self-organizing behavior of block
copolymer systems becomes especially pronounced is the strong
segregation limit, in which monomers of different types segregate
almost completely into non-overlapping regions of space. In the case
of diblock copolymers, the first theory of the strong segregation
limit was proposed by Semenov, who computed the phase coexistence
boundaries for several types of microphases with simple geometries
\cite{semenov85}. His results were essentially corroborated by the
computational studies of self-consistent mean-field models and
extended under a number of simplifying assumptions to include more
exotic phases \cite{matsen94,matsen96jcp,likhtman94,olmsted98}.
Nevertheless, the original Semenov's theory crucially relies on
approximating the Wigner-Seitz cell of the corresponding periodic
structure by a disk of the same area in the case of cylindrical
phases, or a ball of the same volume in the case of spherical
phases. Thus, Semenov's theory neglects the effect of the cell
geometry and, therefore, cannot distinguish between the structures
which are characterized by different unit cell types (as, e.g.,
body-centered cubic versus face-centered cubic lattices of
spheres). This point becomes even more critical when considering the
effects of confinement, since the domain shape must be truly important
in such problems.

In this paper, we propose an extension of the strong segregation theory
for block copolymer systems which includes the effect of geometric
constraints on the chain configurations. Specifically, we investigate
the case of diblock copolymers with strong composition asymmetry, in
which most of the ``unpleasant'' features of the strong segregation
limit \cite{matsen02} are under control, allowing us to concentrate on
essentially the only remaining issue of the effect of the geometry. We
formulate a variational problem that gives the excess energy due to
geometric factors and, in particular, allows to discriminate between
different lattice types with the same unit cell volume in the case of
periodic microstructures. To illustrate the latter point, we
explicitly compute the excess energies for cylindrical phases on three
fundamental two-dimensional lattices and demonstrate that, as
intuitively and experimentally expected, the hexagonal lattice has the
lowest energy.

Consider a system of linear polymer chains consisting of $N$ monomers
of type A bonded covalently with $f N$ monomers of type B, with $f \ll
1$. If each A- and B-monomer has excluded volume $v$, then $f$ is
basically the volume fraction of the B-monomer. We introduce the Flory
interaction parameter $\chi$, the Kuhn statistical length $b$, and the
root-mean-square end-to-end distance $R \simeq b \sqrt{N}$. The
confinement domain (or the Wigner-Seitz cell for periodic structures)
is denoted by $\Omega \subset \mathbb R^3$.

In the strong segregation regime $\chi N \gg 1$ the A- and B-monomers
locally segregate into disjoint subsets $\Omega_A$ and $\Omega_B$ of
$\Omega$, with a sharply defined interface $\Gamma = \partial \Omega_A
\cap \partial \Omega_B$ containing the A-B junctions. Based on this
observation and the Gaussian chain model, Semenov computed the free
energy of the system as a sum of three contributions:
\begin{eqnarray}
  \label{Fsum}
  F = F_\mathrm{interface} + F_\mathrm{core} + F_\mathrm{corona}. 
\end{eqnarray}
Here, $F_\mathrm{interface}$ is the interfacial energy given by
\begin{eqnarray}
  \label{Fint}
  F_\mathrm{interface} = \sigma \int_\Gamma dA, \qquad \sigma =
  \frac{\sigma_0  k_B T \chi^{1/2} b}{v},
\end{eqnarray}
where $\sigma_0$ is a dimensionless parameter of order 1, $k_B$ is the
Boltzmann constant and $T$ is temperature. Next, $F_\mathrm{core}$ is
the energy of the small B-monomer core, assumed to be radially
symmetric \cite{semenov85,likhtman94,matsen02}:
\begin{eqnarray}
  \label{Fcore}
  F_\mathrm{core} = \frac{3 \pi^2 k_B T}{8 v N^2 b^2 f^2}
  \int_{\Omega_B} z^2(\mathbf r) \, d \mathbf r,
\end{eqnarray}
where $z(\mathbf r)$ is the distance from $\mathbf r$ to
$\Gamma$. Finally, $F_\mathrm{corona}$ is the energy of the corona
composed of the A-monomers filling $\Omega_A$. Both calculations of
$F_\mathrm{core}$ and $F_\mathrm{corona}$ rely on the strongly
stretched Gaussian chain assumption. However, the computations differ
crucially, since the surface $\Gamma$, as seen from $\Omega_A$, is
convex, while from the side of $\Omega_B$ it is concave. A parabolic
brush assumption can be used to compute $F_\mathrm{core}$, while one
needs to take into account the exclusion zone to compute
$F_\mathrm{corona}$
\cite{semenov85,ball91,milner91,semenov93,likhtman94,belyi04,matsen02}. In
the original Semenov's theory, the domain $\Omega$ is replaced with a
cylindrical or spherical region of the same volume, and the
Alexander-de Gennes brush assumption \cite{alexander77,degennes76} is
used to compute the energy. This, however, does not affect the leading
order contribution to the energy at small volume fractions, since the
energy is dominated by the singularity near the interface
\cite{semenov85,ball91,semenov93,fredrickson93}.

We will now compute the corona energy, taking also the geometric
effects into account. We start with the self-consistent mean-field
theory in which each polymer molecule is treated as a Gaussian chain
with the position $\mathbf r_0 \in \Gamma$ of the A-B junction
uniformly distributed on the interface (the latter is justified for $f
\ll 1$). Then the corona free energy is (up to an additive constant)
\begin{eqnarray} 
  \label{F} 
  \frac{F_\mathrm{corona}}{k_B T} = -\nu  \int_\Gamma
  \left( \ln 
    \int_{\Omega_A^N} e^{ - H_A / k_B T } \prod_{n = 1}^N d \mathbf
    r_n \right) dA. 
\end{eqnarray}
Here, $\mathbf r_1, \ldots, \mathbf r_N$ denote the positions of the
A-monomers, $\mathbf r_0$ is the position of the A-B junction, $\nu$
is the density of the A-B junctions on $\Gamma$, and $H_A$ is the
Gaussian chain Hamiltonian:
\begin{eqnarray}
  H_A & = & \frac{3 k_B T}{2 b^2} \sum_{n = 1}^N (\mathbf r_n -
  \mathbf r_{n-1})^2 \nonumber \\ 
  & + & \sum_{n = 1}^N \varphi_A(\mathbf r_n) -  \frac{
    \int_{\Omega_A} \varphi_A(\mathbf r) d \mathbf r}{\nu v
    \int_\Gamma dA}, 
\end{eqnarray}
where $\varphi_A$ is the self-consistent field (a Lagrange multiplier)
enforcing the average monomer density to be $v^{-1}$ everywhere in
$\Omega_A$. 

In the strong segregation limit the chains are highly stretched,
i.e. we have $|r_N - r_0| \gg R_g$, so one would naturally want to use
the method of steepest descent to evaluate the integral in
(\ref{F}). However, to proceed further we note a general difficulty
that in the strong segregation limit the integral in (\ref{F}) is not
dominated by the global minimizer of $H_A$, the fluctuations of the
end-point positions $\mathbf r_N$ actively contribute to the free
energy of the chains \cite{matsen02}. Nevertheless, when $\Gamma$ is
convex, as seen from $\Omega_A$, an exclusion zone must form around
$\Omega_B$ which is free of the chain ends
\cite{semenov85,ball91}. Moreover, for $f \ll 1$ this exclusion zone
must occupy most of $\Omega_A$, pushing chain ends close to the cell
boundary. In the case when $\Omega$ is a disk an exact solution to the
problem shows that the exclusion layer extends to the fraction of
$2/\pi \simeq 0.64$ of the disk radius, and in fact the majority of
the chain ends are located within about 6\% of the outer boundary
\cite{ball91}. For a sphere the distribution of chain ends is even
tighter, with the dead layer extending to about 0.76 of the radius,
with the majority of the ends within about 4\% of the outer boundary
\cite{belyi04}. Therefore, for sufficiently small values of $f$ a very
good approximation to the problem with an exclusion zone should be
given by the Alexander-de Gennes brush \cite{alexander77,degennes76},
in which all chain ends are assumed to lie on the outer boundary
$\partial \Omega$ \cite{semenov93}. In the following, we adopt this
approximation to eliminate the need to deal with the precise chain end
statistics. We also note that this assumption is expected to be
asymptotically exact when the free end of the A-chains is capped by
sticky end-groups\cite{sijbesma97,ruokolainen98} or by a short block
of C-monomers which is immiscible with either A- or B-blocks.

\begin{figure*}[t]
  \centering
  \includegraphics[width=12cm]{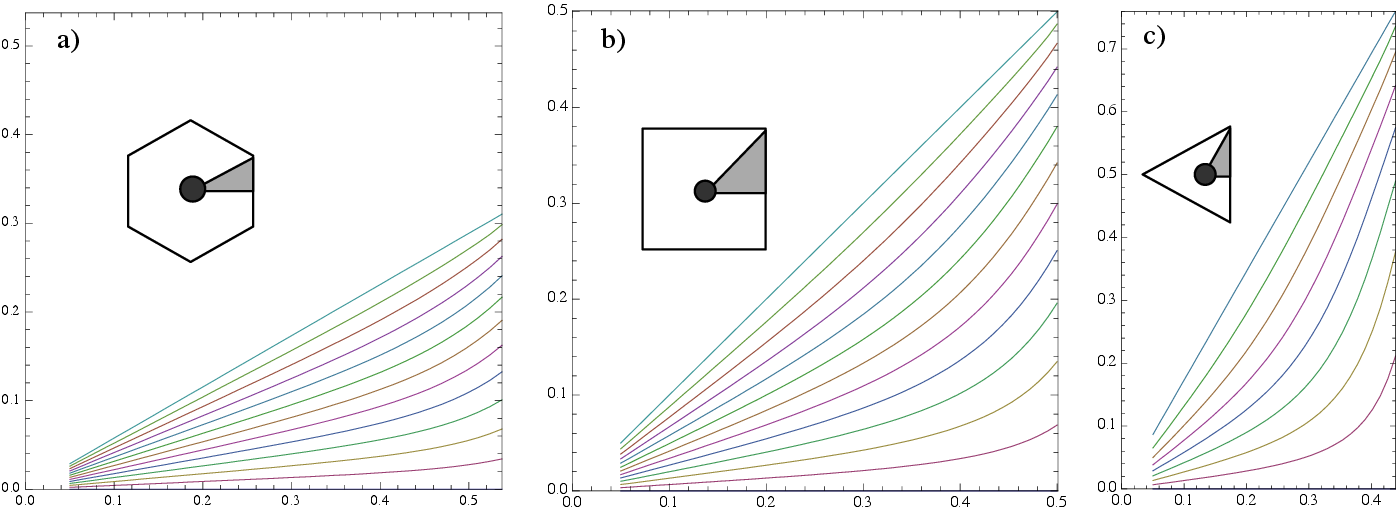}
  \caption{The minimal action paths for the hexagonal (a), square (b),
    and triangular (c) lattices of cylinders. The gray-shaded regions
    indicate the computational domain in each cell. Black circles
    indicate the minority domains.}
  \label{fig:paths}
\end{figure*}

Under the assumption of Alexander-de Gennes brush for the corona, the
integral in (\ref{F}) is dominated by the minimizers of $H_A$ with
$\mathbf r_0 \in \Gamma$ fixed and $\mathbf r_N$ restricted to
$\partial \Omega$:
\begin{eqnarray}
  F_\mathrm{corona} \simeq \nu \int_\Gamma \min_{\{\mathbf r_1,
    \ldots, \mathbf r_{N-1}\}} H_A \, dA(\mathbf r_0). 
\end{eqnarray}
Now, introducing
\begin{eqnarray}
  S = \frac{N b^2}{3 k_B T} \, H_A, \qquad U = - \frac{N^2 b^2}{3
    k_B T} \, \varphi_A,
\end{eqnarray}
and then passing to continuous chains: $\mathbf r_n \simeq \mathbf
r(n/N)$, where $\mathbf r: [0, 1] \to \Omega_A$ are continuous paths,
we can write the corona energy as
\begin{eqnarray}
  F_\mathrm{corona} \simeq \frac{3  k_B T \nu}{2 N b^2}  \int_\Gamma
  \int_0^1 \left| \frac{d \mathbf r}{dt} \right|^2 dt \, dA, 
\end{eqnarray}
where each path $\mathbf r(t)$ minimizes
\begin{eqnarray}
  S = \int_0^1 \biggl\{ \frac12  \left| \frac{d \mathbf r}{dt} 
  \right|^2 -U(\mathbf r) \biggr\} dt
\end{eqnarray}
with fixed endpoints. 

Note that to be in the mechanical equilibrium the chains must come out
normally from the interface $\Gamma$. Then, from the constant monomer
density requirement near $\Gamma$ one can get the initial conditions
for the minimizers $\mathbf r(t)$, given a potential $U(\mathbf r)$
enforcing the constraint:
\begin{eqnarray}
  \label{ic}
  \mathbf r(0) = \mathbf r_0, \qquad \frac{d \mathbf r(0)}{
    dt} = N v \nu \, \mathbf n(\mathbf r_0),
\end{eqnarray}
where $\mathbf n$ is the outward (from $\Omega_B$) normal to $\Gamma$
at $\mathbf r_0 \in \Gamma$. 

We now point out a mechanical analogy, according to which $\mathbf
r(t)$ can be interpreted as the trajectory of a point particle with
unit mass in $\mathbb R^3$ moving under the action of potential energy
$U$. The function $S$ plays the role of the action
\cite{landau1}. Therefore, the equation of motion for $\mathbf r(t)$
becomes simply
\begin{eqnarray}
  \label{mot}
  \frac{d^2 \mathbf r}{dt^2} = -\nabla U(\mathbf r).
\end{eqnarray}
Note that the initial condition in (\ref{ic}) then uniquely determines
the point at which the trajectory $\mathbf r(t)$ hits $\partial
\Omega$.  We can also easily write down the corresponding
Hamilton-Jacobi equation \cite{landau1}:
\begin{eqnarray}
  \label{HJ}
  \frac{d \mathbf r}{d t} = \nabla S(\mathbf r), \qquad \frac12
  |\nabla S|^2 + U = U_0,
\end{eqnarray}
where $U_0$ is a constant which a posteriori turns out to be
independent of the initial point $\mathbf r_0$ of the trajectory that
passes through $\mathbf r$.

Note that we still need to determine the self-consistent field, now
given by $U$, enforcing the constant monomer density constraint, which
in terms of $\mathbf r(t)$ becomes
\begin{eqnarray}
  \label{cons}
  \nu N \int_\Gamma \int_0^1 \delta(\mathbf r - \mathbf r(t)) dt \, dA
  =  v^{-1}, 
\end{eqnarray}
where $\delta(\mathbf r)$ is the three-dimensional delta-function, and
we assumed that the family of trajectories $\mathbf r(t)$ foliates
$\Omega_A$. On the other hand, observe that it is, in fact, sufficient
to find only the action $S$ appearing in (\ref{HJ}). The Hamiltonian
structure of equation of motion (\ref{mot}) imposes certain
restrictions on the possible trajectories $\mathbf r(t)$, in
particular, it forces the dynamics of $\mathbf r$ to be a gradient
flow. In fact, it is easy to see that this gradient flow also has to
be divergence-free. Indeed, consider a tube formed by trajectories
originating on some closed curve in $\Gamma$ enclosing an area $A$,
and write down the total number $M$ of A-monomers contained in the
cylinder between $t = t_0$ and $t = t_0 + \tau$ cross-sections of that
tube. One easily gets $M = \nu N A \tau$, which is clearly independent
of $t_0$. Hence, differentiating this quantity with respect to $t_0$,
we see that the total flow in/out of the cylinder along the
trajectories must equal zero. In view of arbitrariness of $t_0, \tau$
and $A$, we must have $\nabla \cdot (d \mathbf r/dt) = 0$ in
$\Omega_A$.

The arguments above immediately imply that the action $S$ must be a
harmonic function:
\begin{eqnarray}
  \label{harm}
  \Delta S = 0 ~ \mathrm{in} ~ \Omega_A, \qquad \mathbf n \cdot
  \nabla S = N v \nu ~\mathrm{on} ~\Gamma,
\end{eqnarray}
where $\Delta$ is the Laplacian, and we also used (\ref{ic}). On the
other hand, the boundary data on $\partial \Omega$ must be chosen in
an unusual way: {\em every trajectory starting on $\Gamma$ at $t = 0$
  and flowing up the gradient of $S$ must reach $\partial \Omega$ at
  $t = 1$}. This condition can also be reformulated as:
\begin{eqnarray}
  \label{fl}
  \int \frac{|d \mathbf r|}{|\nabla S(\mathbf r)|} = 1 ~~\mathrm{on~
    every ~field ~line~of~} S.
\end{eqnarray}
It is also easy to see from (\ref{cons}) that
\begin{eqnarray}
  N v \nu \int_\Gamma \int_0^1  \left| \frac{d \mathbf r}{dt}
  \right|^2 dt \, dA = \int_{\Omega_A} |\nabla S|^2 \, d \mathbf r. 
\end{eqnarray}
With this, the expression for the corona energy becomes
\begin{eqnarray}
  \label{dir}
  F_\mathrm{corona} \simeq \frac{3  k_B T}{2 v N^2 b^2}  \, D[S], \quad
  D[S] =  \int_{\Omega_A} |\nabla S|^2 \, d \mathbf r. 
\end{eqnarray}

Let us note that existence and uniqueness of solutions to the proposed
problem is not guaranteed a priori for any given $\Omega$ and
$\Gamma$. In view of (\ref{dir}), one should, in fact, be interested
in {\em minimizers} of $D[S]$ which also satisfy the condition in
(\ref{fl}). Notice that the location of $\Gamma$ relative to $\Omega$
is also part of the minimization problem. A radial solution
(presumably, the unique minimizer) trivially exists when $\Omega_B$
and $\Omega$ are concentric balls. One would then expect from
perturbative considerations that this solution should persist when
$\partial \Omega$ is slightly distorted away from a sphere.

In view of the assumption $f \ll 1$, the solution of (\ref{harm})
coincides to the leading order with that of
\begin{eqnarray}
  \Delta S = |\Omega| \delta(\mathbf r).
\end{eqnarray}
Indeed, if $\Omega_B$ is a ball of radius $R$ centered at the origin,
then $4 \pi R^2 \nu N v = |\Omega|$ to the leading order in $f$, where
$|\Omega|$ denotes the volume of $\Omega$ (same argument applies in
two dimensions). From this, one can see that since $S$ should behave
as the free space Green's function of the Laplacian near the origin,
$D[S]$ will diverge as $R \to 0$. The leading order singular term will
only depend on $|\Omega|$ and not the shape of $\Omega$ and is
precisely what was calculated by Semenov for the corona energy
\cite{semenov85}. On the other hand, for small but finite values of
$R$ the solution will also contain an {\em excess energy} associated
with the geometry of $\Omega$.

We applied our variational procedure to compute the excess energy in
the case of two-dimensional hexagonal, square, and triangular lattices
of straight cylinders $\Omega_B$ of radius $R$ and unit height. We
note first that the exact solution of the problem in a coaxial
cylinder of the same volume $|\Omega|$ gives straightforwardly
\begin{eqnarray}
  D_0 = \frac{|\Omega|^2}{4 \pi} \ln f^{-1}. 
\end{eqnarray}
To compute the excess energy $D - D_0$ for Wigner-Seitz cells
corresponding to the considered lattices, we implemented a finite
element-based minimization algorithm to find minimizers of $D$
satisfying (\ref{fl}) \cite{mnog}. The minimizing trajectories in
cells whose area is normalized to unity are presented in
Fig. \ref{fig:paths}. From dimensional arguments, we find that for $f
\ll 1$ we have
\begin{eqnarray}
  D - D_0 \simeq C_\Omega |\Omega|^2,
\end{eqnarray}
where the dimensionless constant $C_\Omega$ depends on the geometry of
the cell only. Numerically, we found $C_\mathrm{hex} \simeq 0.00922$,
$C_\mathrm{sq} \simeq 0.0453$, and $C_\mathrm{triang} \simeq 0.179$
for the hexagonal, square, and triangular lattices, respectively. Note
that for fixed values of $R$ and $f$ both the interfacial energy
$F_\mathrm{interface}$, the core energy $F_\mathrm{core}$, and the
leading-order corona energy $F_\mathrm{corona}$ obtained from
(\ref{dir}) with $D$ replaced by $D_0$ are the same. Therefore, to
compare the energies of different geometric arrangements of the
B-domains, one needs to compare the excess energies. From our
calculation above we can immediately conclude that among the
considered types of lattices of cylinders with the same radius $R$ and
volume fraction $f$ the hexagonal lattice is the most energetically
favorable in the limit $f \to 0$, an intuitively expected result which
is put on a rigorous footing by our computations. Let us point out
that our approach should also be applicable to spherical phases to
help identify the minimizer among different types of three-dimensional
lattices. We note that the answer to this question in the strong
segregation limit lies beyond the scope of Semenov's theory
\cite{semenov85} and its extensions \cite{semenov93,likhtman94}. Let
us also point out that the method of Refs. \cite{likhtman94,olmsted98}
cannot be applied here, since it ignores the effect of the exclusion
zone.

Let us note that our calculation is akin to the one performed by
Fredrickson \cite{fredrickson93}, who estimated the excess energy due
to geometric factors for a hexagonal Wigner-Seitz cell in the strong
segregation limit. Fredrickson used linear elasticity and the
Alexander-de Gennes assumption to study the extra contribution to the
elastic energy of the corona due to chain distortions. His result,
however, differs from ours quantitatively. In particular, we find the
excess energy obtained by us is greater than the one obtained by
Fredrickson by a factor of 1.5. We attribute this discrepancy to
strong chain distortions, which invalidate the linear elasticity
approximation. Thus, at small $f$ the excess energy due to geometry of
the Wigner-Seitz cell may have a larger contribution than previously
expected.

To conclude, we have developed a variational characterization of the
leading geometric corrections to the Semenov's strong segregation
theory in the case of strong composition asymmetries. Our theory thus
should be able to account for the effect of the confinement geometry
on microstructures consisting of small droplets of the minority
species and, in particular, help identify the equilibrium lattice
configurations of these droplets, as was explicitly demonstrated in
the case of cylindrical phases. Perhaps more importantly, our theory
provides a new way to study questions of metastability and instability
of nonequilibrium copolymer microstructures under external
perturbations \cite{ohta86,m1:prl97,m:pre02,matsen06}.

We wish to thank G. Fredrickson for valuable comments. The work of
C. G.-C. was supported by NSF DMS-0505738 grant. C.B.M., M.N., and
G.O. gratefully acknowledge support by GNAMPA.

\bibliography{../nonlin,../mura,../stat}

\end{document}